# Theoretical developments in heavy nuclei

## Jacek Dobaczewski

*Institute of Theoretical Physics, Warsaw University, ul. Hoża 69, PL-00681, Warsaw, Poland*

**Abstract.** The present-day nuclear structure theory exhibits a great degree of synergy with respect to methods that are used to describe various phenomena in heavy nuclear systems. From few-body methods, through the shell model to mean-field approaches, the bridges are being built between different ways of describing the stable as well as the most exotic nuclei. In the present talk, I give a review of several selected subjects that are currently at the fore front of new developments in this domain of nuclear science.

## INTRODUCTION

Heavy atomic nuclei belong to the class of many-body systems that are too complex for a direct description from first principles, and too small for a quantitative statistical approach. Quantum effects, strong interactions, and repulsive core are common features of several physical systems, like nuclei, metal clusters and grains, and atomic clouds, that are at present intensively studied.

The fact that nucleon-nucleon (NN) interaction is a long-range remnant of the QCD interactions between quarks and gluons coupled to colorless states, and hence is not known in a closed form, adds all the more to the complexity of nuclear systems. Although on-shell properties of the NN interactions in the vacuum are phenomenologically very well known and parametrized, see, e.g., Ref. [1], the off-shell properties are not. Nevertheless, light nuclei (up to $A$=10 nucleons) can be with a large success described by direct methods based on phenomenological NN potentials [2, 3]. However, much work is being performed, and even more is needed, to achieve a more complete state of knowledge of the NN and NNN interactions.

A direct description of heavy nuclei, based on the NN interactions and leading to a complete wave function, is neither feasible nor, in fact, sensible. Only a limited number of specific states of heavy nuclei, mostly at low energies, can be studied in experiment. These states do not depend on the whole complexity of the NN interactions, and do not probe the whole $A$-body Hilbert space. Therefore, a great deal of effort and success in nuclear structure physics has been devoted to deriving effective interactions that would describe main features of nuclei within a restricted space. Due to applications of the effective field theory and renormalization group techniques, such derivations are recently gaining dramatically new momentum [4, 5].

Shell-model calculations that use derived effective interactions are by now possible for systems with $A$∼12 nucleons [6]. These methods look very promising, and can probably be applied to still heavier systems. It is extremely encouraging that similar shell-model calculations using phenomenological effective interactions fitted to data, are very successful in describing properties of all nuclei with up to about $A$∼50 particles [7]. In special cases such methods can be even used in much heavier systems. This shows that basic low-energy properties of nuclei are indeed governed by relatively simple two-body effective interactions acting in manageable Hilbert spaces, and gives us hope that derived effective interactions may become a link between the NN forces and the present-day phenomenology in heavy nuclei.

Significant progress was achieved in recent years in a phenomenological mean-field description of a multitude of nuclear states and phenomena. Although detailed formulations vary, most mean-field approaches are presently based on the energy density formalism, either in a relativistic or non-relativistic flavor. On the other hand, much too little work has been done to derive the energy density functionals from the underlying NN forces and/or effective interactions. This direction of research is far behind the analogous efforts and successes achieved in molecular physics, and will probably see a substantial increase of activity in the coming years.

In general, mean-field methods correctly reproduce the main features of structure of heavy nuclei. Very often, however, this is not sufficient, and a higher degree of precision is requested from theoretical descriptions in order to understand fine-scale nuclear phenomena. In particular, the limits of nuclear binding depend on a detailed balance between the bulk, surface, and pairing effects, see, e.g., Ref. [8]. Although a large progress has recently been achieved [9], a microscopic description of nuclear binding energies within a $100\,\text{keV}$ precision is, for the moment, an unreachable goal. Similarly, detailed properties of rotational bands give us extremely precise information on the structure of microscopic states, and are fairly well understood on the phenomenological level, whereas a link to the corresponding terms in the energy functional is still unclear.

In the present talk, a number of recent specific achievements in the theoretical studies of heavy nuclei are briefly discussed. For example, deformations and halos in heavy neutron-drip nuclei, as well as widths of one-body resonances, were calculated with the coupling of bound and continuum states taken into account. A great progress in understanding proton-neutron correlations in $N{\sim}Z$ nuclei was achieved. New kind of collective motion, the magnetic rotation, was predicted and found in experiment. Origins of the pseudospin symmetry are now being traced back to simple relativistic single-particle effects. Self-consistent calculations of superheavy nuclei gave us a new insight into the shell structure in the presence of a very strong Coulomb field. Since it would be impossible to give here a full bibliography of the discussed subjects, only the recent works are usually cited, and more references can be found there. There are also many other subject that cannot be presented here because of limited space; some of them are covered in other plenary talks [7, 10, 11] during this conference. The reader may also wish to consult the contents of nuclear-structure talks that have been given during the previous INPC'98 conference [12], where the three-years-ago status of the field was presented.

## SHELL MODEL EMBEDDED IN THE CONTINUUM

Theoretical description of properties of weakly bound nuclei requires new methods that take into account the continuum phase space explicitly. Traditional shell-model approaches use well-bound single-particle states, most frequently the harmonic oscillator (HO) basis, to define the Fock space of many-nucleon configurations. Within such a space, particle emission threshold is completely absent, and a large fraction of obtained excited states (even all of them in case of weakly bound nuclei) have excitation energies above the threshold and, at the same time, localized wave functions. This inconsistency precludes by construction any possibility to describe decay widths of excited states, and moreover, introduces an unphysical cut-off in the physically important low-energy phase space.

Improvements upon this unsatisfactory situation have been proposed and defined since many years, see, e.g., Ref. [13], however, only very recently practical applications were realized. The shell model embedded in the continuum (SMEC) [14, 15] combines the state-of-the-art shell model methods and interactions with the exact treatment of the one-body continuum. It means that the standard space of shell-model states of an $A$-particle system is enlarged by the similar shell-model states of the $(A-1)$-body system coupled with single-particle continuum states. The one-body continuum is appropriately orthogonalized with respect to the space of bound single-particle states. In the SMEC, the biggest unknown factor is the interaction between the shell-model and continuum states (the simplest contact forces have been used up to now), and this aspect certainly requires further analysis. The known nucleon-nucleus potentials may not be directly applicable, because interactions with all states of an $(A-1)$-body system have, in principle, to be taken into account.

Results of calculations performed for the $p+^{16}$O reactions are presented in Fig. 1. The obtained quality of description is indeed very satisfactory. One is able to correctly reproduce the total and differential cross-sections as well as the widths of low-excited unbound states. The model may fail in predicting the widths of states for which the two- or three-body continuum becomes available, and the inclusion of these channels is probably one of the most challenging future extensions of the model.

## TWO-PROTON DECAY

A search for the two-proton ground-state radioactivity has a long history, however, apart from the well-known three-body ($\alpha pp$) decay of $^{6}$Be, such decay mode has not yet been experimentally discovered. One hopes that simultaneous emission of two protons may give us invaluable information about the two-particle ground-state correlations in nuclei. Such an emission explores a different phase space than the standard two-particle transfer experiments which were used to study pairing correlations in nuclei. Indeed, in the emission process the two-protons may leave the nucleus in various final states, ranging from a quasi-bound diproton to anticorrelated protons emitted in opposite directions, while in transfer reactions they are absorbed from a given initial state.

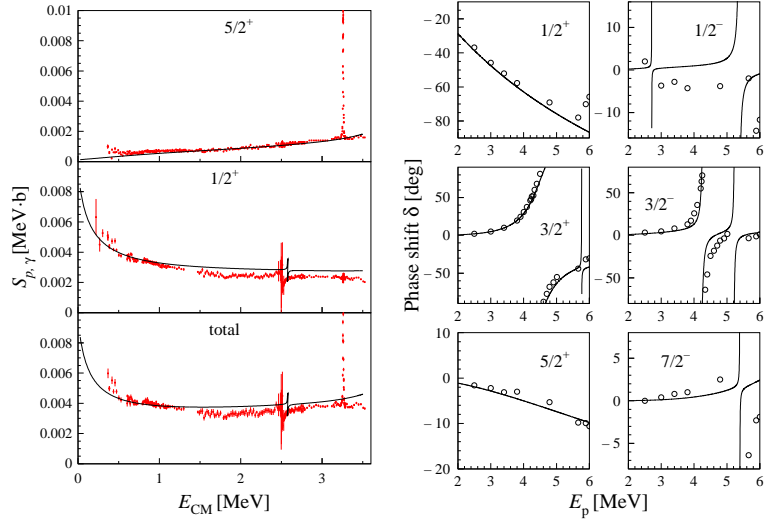

**FIGURE 1.** The astrophysical *S*-factors (left) and phase-shifts (right) for the *p*+$^{16}$O reactions, calculated within the SMEC and compared with experimental data. (From Ref. [15]. Reprinted from Physics Letters B, Vol 488, 2000, Page 75, K. Bennaceur *et al*, "Shell model description of...", Copyright (2000), with permission from Elsevier Science.)

The simultaneous two-proton emission, whenever it is energetically possible, i.e., only beyond the proton drip line, must compete with two other decay modes which are usually much more probable, namely, the one-proton emission and β decay. Therefore, one has to look for nuclei in which one-proton decay is energetically impossible (which limits the selection to even-*Z* elements), and where the two-proton decay energy is sufficiently large. During the emission process, the two protons must tunnel through a wide Coulomb barrier, typically traversing the classically forbidden region of even up to about 100 fm. Hence, the emission probability very strongly depends on the available $Q_{2p}$ value. For small $Q_{2p}$ values, β decay is more probable, while for the large ones the life time of the given nucleus is very short. Therefore, the window of opportunity to observe the two-proton decay is fairly small. Nevertheless, several candidate nuclei were already identified ($^{19}$Mg, $^{42}$Cr, $^{45}$Fe, $^{48,49}$Ni,...), and intense searches of this new decay mode are presently conducted, see e.g. Ref. [16].

Two-proton emissions from excited states was already observed in different configurations, such as the decays of analog states, e.g. from $^{31}$Cl (see Ref. [17] and references cited therein), or resonances, see Ref. [18] where the decay of $^{18}$Ne was studied. All these experiments encounter difficulties in extracting information about the decay process from observed data. Indeed, in order to differentiate various two-proton decay scenarios one has to perform the full three-body calculations, where the two protons and the daughter nucleus are allowed to develop arbitrary correlations, and the full three-body wave function is obtained. The simple diproton emission can be calculated very easily within the WKB approximation (see Ref. [19] and references to earlier papers cited therein). However, the three-body calculations with Coulomb interaction are much more difficult, and have become available only very recently [20, 21].

Such calculations reproduce the experimental width of the $^6$Be decay very well. Examples of results obtained for $^{19}$Mg and $^{48}$Ni are shown in the left panel of Fig. 2 [20]. It can be seen that the widths based on the diproton hypothesis and those obtained by assuming a direct decay to continuum from the ℓ=0 state are very similar. The later widths strongly depend on ℓ, and are much smaller for values corresponding to relevant shell-model orbitals occupied at a given number of protons. Results obtained by the full three-body calculations are by about three orders of magnitude smaller than those obtained for the ℓ=0 diproton emission. This shows that the final-state proton-proton interaction is not sufficient to maintain this pair of particles in a correlated state during the emission process. Since the full three-body wave function of the complete system is obtained, one can determine and compare with experiment not only the life time, but also all relevant correlation observables, e.g., the relative angles and energy correlations. This is crucial for the attempts to determine initial-state correlations from experiment. However, reliable conclusions about the proton-proton ground-state correlations can be obtained only from calculations that release the assumption of the inert core, and take the effects of core configuration mixing into account. In this sense, the method used in Refs. [20, 21] is complementary to that developed in the SMEC model discussed in the previous section. The former one treats the two-body continuum exactly but assumes an inert core, while the latter one neglects the two-body continuum but uses the full shell-model space for the core. Clearly, a combined approach is very much called for.

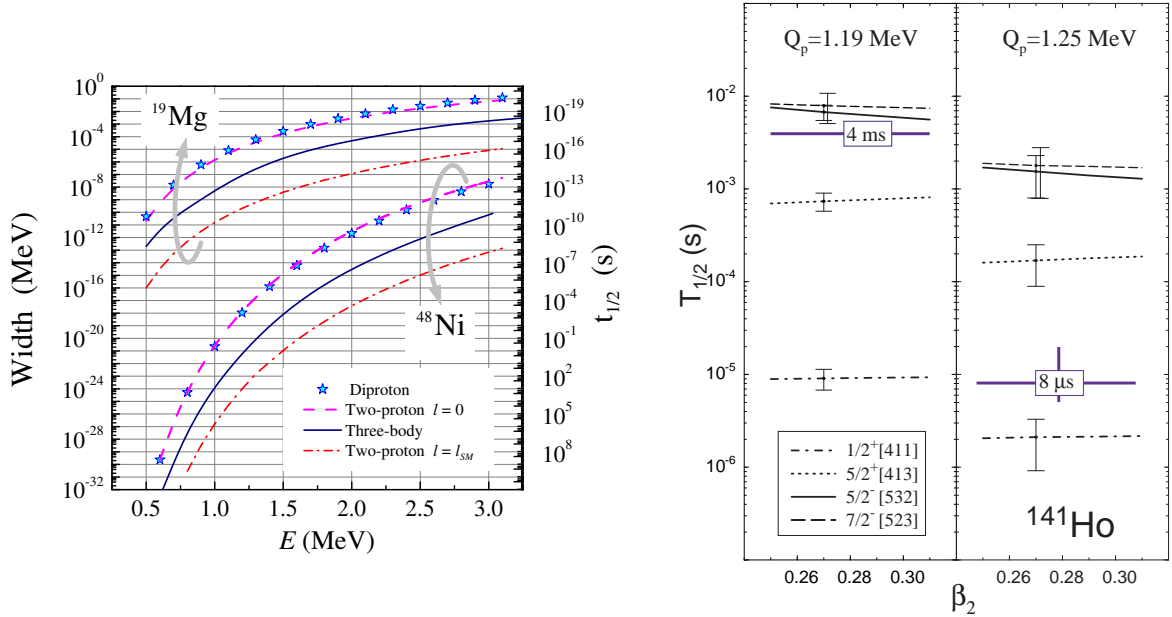

**FIGURE 2.** Left: Width and life time as functions of the ground-state energies of $^{19}$Mg and $^{48}$Ni counted with respect to the two-proton decay threshold. Results of the three-body calculations (solid lines) are compared with those obtained within the diproton and two-proton approximations. From Ref. [20]. Right: Calculated life times of the ground-state ($Q_p$=1.19 MeV) and isomer ($Q_p$=1.25 MeV) proton emission from $^{141}$Ho. Different lines correspond to various Nilsson orbitals. Measured values indicate that the ground-state and isomer emission occurs from the $7/2^-$[523] and $1/2^+$[411] Nilsson orbitals, respectively. From Ref. [22].

# DEFORMED PROTON EMITTERS

A large number of odd-$Z$ elements has isotopes that are sufficiently proton-rich to be unbound with respect to the ground-state proton emission. To date, about 40 such proton emitters are experimentally known [23, 24]. Proton emission from spherical nuclei is well understood in terms of the simplest WKB barrier penetration. Such an approach allows for a consistent identification of the single-particle orbit from which the emission occurs, because the barrier is strongly influenced by the centrifugal component on top of the standard Coulomb barrier. The success is undoubtedly due to the fact that no pre-formation factor has to be here taken into account, contrary to the physical situation pertaining to the $\alpha$-particle emission. The precision of the theoretical description allows to determine the proton spectroscopic factors by comparing the calculated and measured proton-decay life-times, see Refs. [25, 26] and references cited therein.

Recently, the focus of studying proton emitters, both in experiment and in theory, is on investigating such a process when it occurs from deformed nuclei. Here, one has to reconcile the picture of the barrier penetration, pertaining to the intrinsic frame (and the barrier height depending on the direction with respect to the shape principal axis), with the necessity of correctly describing the angular momentum content of orbitals, which defines the corresponding centrifugal barrier. In recent years, a number of theoretical tools have been developed by several groups in order to tackle the problem, and again it seems that the theoretical description gives the full account of experimental data. The method of choice here is the coupled-channel approach that takes into account the structure of collective states in the daughter nucleus [27, 28, 29] (the so-called non-adiabatic approach). Within the strong-coupling adiabatic approach the calculations were also performed by using the deformed resonance states [30, 31], Green function methods [32], reaction theory [33], and time-dependent transmission calculations [34].

The challenge of achieving the best possible description of deformed proton emitters lies in a possibility of determining with a large precision the Nilsson quantum numbers of deformed single-particle states in nuclei which cannot be accessed by other spectroscopic methods. An example of such analysis is shown in the right panel of Fig. 2 [22]. Measured life times of protons emitted from the ground and isomeric states [22, 35] provide a stringent test on the type of single-particle deformed orbitals occupied in [141]Ho. Although the branching ratios to excited states in the daughter nucleus [140]Dy are not yet known, the yrast cascade up to $8^+$ was recently identified in this nucleus from a decay of a $20\,\mu s$ isomer [36]. On the other hand, the fine structure was already observed in the proton decay of [131]Eu to the $0^+$ and $2^+$ states in [130]Sm, which allowed for an unambiguous determination of the corresponding Nilsson orbital in [131]Eu [37, 38, 27]. Similar methods were also used for a study of odd-$N$ proton emitting nuclei [39], where, in principle, the proton-neutron interaction effects could be studied. A possibility to perform a much more difficult experiment on the proton emission from oriented nuclei was discussed in Ref. [40].

## HALOS IN HEAVY NUCLEI

A fascinating phenomenon of increased radial sizes of nuclei far from stability [41, 42, 43, 44] was very intensely studied in recent years. In extreme situations, when nuclei consist of about thrice more neutrons than protons, the outer weakly bound neutrons may form halos of particle densities extending to rather large distances. For example, the flag-ship nucleus [11]Li exhibiting this kind of structure has the measured [45] root-mean-square (rms) radius of $R_{\rm rms}$=3.27(24) fm, while in [9]Li one has $R_{\rm rms}$=2.43(2) fm. Supposing that in [11]Li the size of the "core" subsystem of 9 particles is the same as that of [9]Li, one obtains the rms radius of the two-neutron subsystem equal to $\sqrt{(11*3.27^2 - 9*2.43^2)/2}$=5.67 fm. Therefore, the outer two-neutrons occupy a volume that has the size similar to that of [208]Pb, which is a nucleus with about 20 times more particles!

Sizes of several light neutron-rich nuclei were recently measured via the total interaction cross sections, see Refs. [46, 47, 48]. Although such measurements rely on a number of theoretical assumptions pertaining to the reaction mechanism, the increase of size of nuclei near the drip lines seems to be fairly well established. The question of increased sizes of medium heavy and heavy neutron-rich nuclei could be so far addressed only in theoretical calculations within mean-field models. The decisive role played by pairing correlations in establishing asymptotic density distributions has been recognized since a long time [49, 50], and several calculations based on the Bogoliubov approach were already performed, both in the non-relativistic [51, 52, 53] and in the relativistic [54, 55, 56, 57] framework.

The neutron rms radius constitutes only an indicative general measure of the neutron density distribution. In fact, three different physical properties of the density distribution can be singled out that strongly influence the rms radius, namely, (i) the global size of the nucleus, (ii) the surface diffuseness, and (iii) the rate of decrease of density in the asymptotic region. These three properties are almost directly related to (i) the number of nucleons, (ii) the surface tension or surface interaction terms, and (iii) the particle binding energy or the Fermi energy, respectively. The halo phenomena are obviously related to properties of particle distributions in the asymptotic region. One should keep in mind, however, that we cannot expect nuclear halos to correspond to what one might infer from the every-day-life meaning of the term. The nuclear density distributions always decrease when the distance from the center of nucleus increases, and by the halo we can only mean an unusually slow rate of such a decrease.

In order to disentangle the three effects mentioned above, one has to use a sufficiently rich model of the neutron distribution. For example, the density distributions obtained by filling with particles the harmonic-oscillator (HO) potential up to a given principal HO number cannot serve this purpose, because they are determined by a single parameter, the HO frequency $\hbar\omega_0$. Therefore, neither the surface diffuseness nor the asymptotic decrease rate can be properly described, even if $\hbar\omega_0$ is adjusted to reproduce the overall size of the nucleus. Similarly, the Fermi (or the Woods-Saxon) type of shape,

$$\rho(r) = \frac{\rho_0}{1 + \exp[(r - R_0)/a]} \quad, \tag{1}$$

can reproduce the overall size and surface diffuseness by an adjustment of parameters $R_0$ and $a$, respectively. (See Ref. [58] for a recent application of this form to an analysis of experimental data.) However, here the asymptotic behavior, $\rho(r) \simeq \rho_0 \exp\{-r/a\}$, is dictated by the surface diffuseness $a$, and not by the Fermi energy. In order to describe this latter physical feature, that is crucially important in weakly bound systems, even more a complicated model distributions must be used. For example, the third parameter can be introduced in a very simple way by rising the denominator in Eq. (1) to some power $\gamma$. In this case, the surface diffuseness and asymptotic decrease rate can be independently adjusted by varying $a$ and $\gamma$. Another parameterization achieving similar goal was recently analyzed in Ref. [59].

Density distributions obtained within microscopic self-consistent approaches are obviously able to independently model all three physical effects described above. Here, the quantitative reproduction of the nuclear size, surface diffuseness, and asymptotic decrease rate depends on properties of the underlying interactions, while the density distributions may in principle assume unrestricted forms. By employing the classic Helm model distributions [60, 61, 62], a method to analyze microscopic distributions was recently devised [53]. A comparison of Fourier transforms (form factors) of the Helm and microscopic densities allows to define the diffraction radius $R_0$ and the surface thickness $\sigma$ from the positions of the first zero and first maximum of the form factor, respectively. These characteristics of the density are entirely independent of the asymptotic decrease rate, and hence they define a suitable Helm rms reference radius,

$$R_{\mathrm{rms}}^{(\mathrm{H})} = \sqrt{\frac{3}{5}\left(R_0^2 + 5\sigma^2\right)}. \tag{2}$$

An increase of the rms radius with respect to the Helm rms reference radius can now be used as a signal for the appearance of the halo structure, and hence the size of the halo was in [53] defined as the difference $\delta R_{\mathrm{halo}} = \sqrt{\frac{5}{3}}(R_{\mathrm{rms}} - R_{\mathrm{rms}}^{(\mathrm{H})})$. Such a halo parameter allows us to discuss and compare the results obtained within different models, and puts a definition of the halo phenomenon on quantitative grounds.

In Ref. [53] it was shown that in the mean-field description of medium heavy and heavy nuclei, the neutron halos appear gradually when the neutron numbers increase beyond the neutron magic numbers. The halos of different sizes can be obtained by changing the parameterizations of nuclear effective interactions, and in particular those of the pairing force [63]. Results of calculations presented in the left panel of Fig. 3 indicate that the sizes of halos at the two-neutron drip line are of the order of 1 fm throughout the mass table. Fluctuations that are observed in function of the mass number are correlated with the values of the neutron Fermi energy $\lambda_n$. However, no correlation is observed with the positions of the Fermi energy with respect to the low-$\ell$ ($s$ and $p$) single-particle orbitals. This latter fact reflects the so-called pairing anti-halo effect [64].

Indeed, the pairing correlations strongly modify the asymptotic properties of single-particle densities. In the extreme single-particle picture, the single-particle $\ell=0$ wave functions behave asymptotically as $\psi(r) \simeq \exp(-\kappa r)/r$, with the decrease rate of $\kappa = \sqrt{-2m\varepsilon/\hbar^2}$ given by the single-particle energy $\varepsilon$. Therefore, with $\varepsilon \to 0$, these wave functions become infinitely flat, and the corresponding rms radii, and the halo sizes, become infinite. Nothing of the sort is observed in the Hartree-Fock-Bogoliubov (HFB) calculations, because here the lower components of the quasiparticle wave functions vanish exponentially with the decrease rate of $\kappa = \sqrt{2m(E - \lambda)/\hbar^2}$ given by the quasiparticle energy $E$ [49, 50]. Since the paring correlations do not vanish at drip lines, the nonzero values of $E$ prevent $\kappa$ from vanishing, even in the limit of zero binding given by $\lambda=0$. As the result, one of the conditions generally thought to be a prerequisite for the halo structures [65], i.e., the presence of the low-$\ell$ orbitals, does not apply to paired even-$N$ systems. On the other hand, this condition does apply to odd-$N$ systems, because the blocked HFB states correspond to the quasiparticle wave functions that vanish with the decrease rate of $\kappa = \sqrt{-2m(E + \lambda)/\hbar^2}$, which goes to zero at the one-neutron drip line given by $E + \lambda=0$.

The fact that the neutron densities of even-$N$ systems decrease with a non-zero rate when approaching the drip line is well born out in all self-consistent Bogoliubov calculations. An example obtained within the relativistic mean field (RMF) approach [54] is shown in the right panel of Fig. 3. Results for all bound even-$N$ sodium isotopes were calculated in space coordinates, and the resulting matter density distributions clearly show a common non-zero decrease rate when approaching the neutron drip line, which simply illustrates the pairing anti-halo effect. This is especially strongly accentuated in the RMF+HB method which gives a cluster of nearly degenerate states close to the Fermi surface [54], which results in strong pairing correlations near the neutron drip line. Exactly the same situation occurs in the zirconium isotopes studied in Ref. [55].

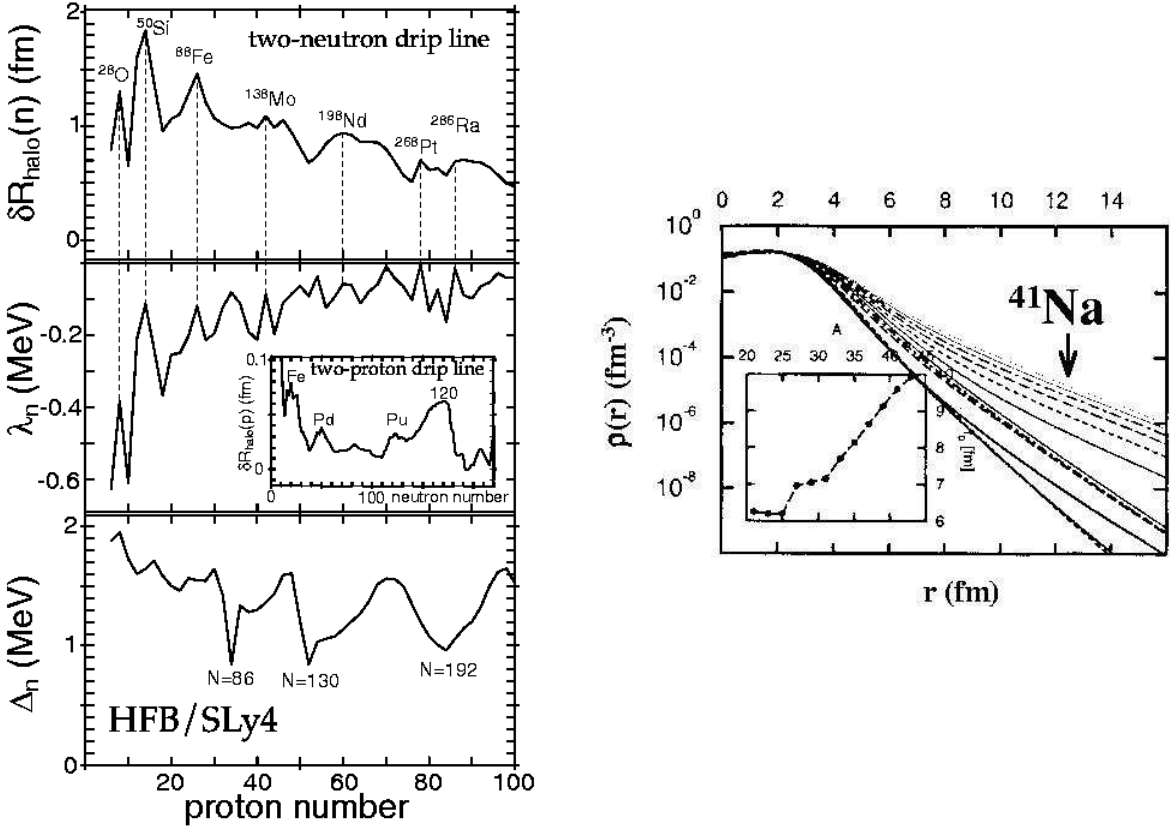

**FIGURE 3.** Left panel: neutron halo parameters (top), neutron Fermi energies (middle), and neutron pairing gaps (bottom) calculated in the HFB/SLy4 model for the two-neutron drip-line even-even nuclei (i.e., the heaviest even-even isotopes which are predicted to be two-neutron bound). From Ref. [53]. Right panel: matter density distributions calculated in sodium isotopes within the RMF+HB approach. Near the neutron drip line ($^{41}$Na) the matter density at large distances is determined by the neutron density. (From Ref. [54]. Reprinted from Physics Letters B, Vol 419, 1998, Page 1, J. Meng *et al*, "The proton and neutron distributions in...", Copyright (1998), with permission from Elsevier Science.)

## MASS PREDICTIONS FAR FROM STABILITY

Changes of the shell structure, that are expected in nuclei far from stability, are one of the most interesting subjects to be studied in the future exotic-beam facilities [66, 67, 68]. In light nuclei, such changes are already observed at N=8, 20, and 28 [11]. They are manifested by modifications of nuclear binding energies, that can be interpreted in terms of an appearance of new shells (e.g., N=16 [69]), or as an unexpected collectivity observed in semi-magic nuclei, e.g., in $^{32}$Mg [70]. In light nuclei, such changes can be attributed to shifts of single-particle levels induced by selected channels of the shell-model interactions (see, e.g., Ref. [71]), or to the shape coexistence phenomena related to intruder configurations, as studied either in the shell model [7] or in the mean-field approach [72, 73, 74]. Both types of effects are strongly connected; indeed, smaller shell gaps in nuclei far from stability may facilitate appearance of intruder configurations at low energies, or polarization induced by the configuration mixing may decrease the shell gaps. Therefore, apart from phenomenological ideas explaining the experimental data, more work is here needed if one wants to discuss the question of whether the shifted levels or intruder states are primary, secondary, or equivalent explanations of the observed changes.

In heavier nuclei, predicted shell modifications can be due to the effects of an increasing surface diffuseness [76], changes in the spin-orbit coupling [77, 78], or influence of pairing correlations [76, 63, 79]. Calculated shell quenching effect is illustrated in Fig. 4 for the nuclei along the N=82 isotonic chain. The left panel presents the spherical two-neutron separation energies $S_{2N}$ which show a conspicuous shell gap between the values corresponding to N=80,82 and N=84,86. When going towards the neutron drip line (i.e., in the direction of decreasing proton number Z), the shell gap decreases, and disappears at the two-neutron drip line (i.e., when the values of two-neutron separation energies go to zero). The shell quenching effect is a generic result obtained in all self-consistent mean-field calculations based on effective two-body interactions. However, the exact magnitude of the effect does strongly depend on what kind of interaction is used. In particular, interactions that give a small effective mass tend to overestimate the shell gaps in stable nuclei, and hence the shell quenching effects are not strong enough to significantly decrease these gaps near the drip lines. Moreover, the type of the pairing interactions used in these calculations crucially influences the degree of the shell quenching [79]. At the moment, our knowledge of the effective interactions is insufficient for reliable predictions of the shell quenching far from stability, and the only method to proceed further seems to be the use of future measurements to fit better interactions.

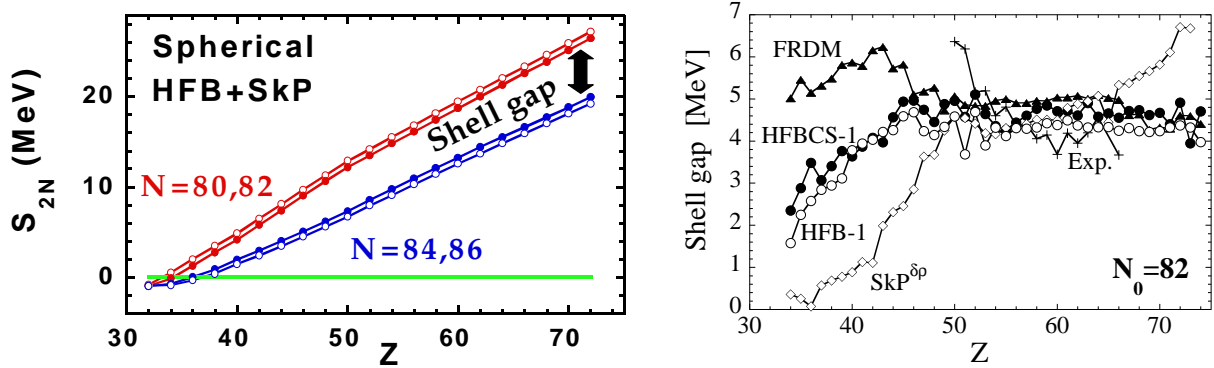

**FIGURE 4.** Two-neutron separation energies (left) and shell gaps (right, from Ref. [75], reprinted from Nuclear Physics A, M. Samyn *et al*, "A Hartree-Fock-Bogoliubov mass formula", in press, Copyright (2002), with permission from Elsevier Science. ) for nuclei around $N$=82 as functions of the proton number $Z$.

The best to date effective interactions fitted to nuclear masses were obtained in Refs. [9, 75]. Depending on the method to treat the pairing correlations, BCS [9] or HFB [75], the authors obtained the rms deviations between the calculated and measured binding energies equal to 0.738 and 0.766 MeV, respectively. The quality of the description is almost identical to that (0.669 MeV) obtained within the microscopic-macroscopic FRDM method based on a phenomenological mean-field potential [80]. Needless to say that masses of unknown nuclei, which are obtained by extrapolating either effective interactions or phenomenological mean-fields, are markedly different. This is illustrated in the right panel of Fig. 4, which shows the shell gaps at $N$=82 defined by $S_{2N}(N$=82$)-S_{2N}(N$=84$)$. Interactions that were fitted to masses give a weak shell quenching as compared the that obtained for the SkP interaction [50], while the FRDM results show no quenching at all. Obviously, there is still a fair degree of arbitrariness in the values of parameters of the effective interactions, and at the moment it is not clear how well these parameters are fixed by fits to masses or other nuclear properties.

Calculations of Refs. [9, 75] were performed by expanding the single-particle wave functions in the harmonic-oscillator (HO) eigenstates, and therefore, they are characterized by unphysical asymptotic properties of the particle and pairing density distributions. It is known [81] that the convergence of results in function of the HO basis size can be extremely slow, and the stability of results tested by increasing the basis by adding one or two HO shells can be misleading. Moreover, the asymptotic properties of densities are essential for a correct description of the pairing channel [51]. It still remains to be investigated to which extend these deficiencies of the method are important for determining masses of weakly bound nuclei. This question can be studied by using methods based on the transformed harmonic oscillator (THO) basis [82, 83, 84] in which the correct asymptotic behavior is obtained by a proper point-like transformation of the basis wave functions.

Preliminary results obtained within such an approach are shown in Fig. 5 [79]. Deformed calculations were performed for a standard Skyrme effective interaction SLy4 [85], and for contact pairing interaction that is intermediate between the volume and surface force [86]. The figure shows the two-neutron separation energies calculated for all even-even nuclei that are bound with respect to two-particle emission, i.e., those that have negative neutron and proton HFB Fermi energies. The results show a very interesting phenomenon occurring at the neutron drip line, namely, *negative* two-neutron separation energies are sometimes obtained for particle-stable nuclei. This apparently contradictory result illustrates the fact that the HFB particle stability is defined with respect to a *fixed* configuration, while the two-neutron separation energies can involve binding energies of neighboring nuclei that may have *different* configurations, e.g., different shapes. Hence, at the neutron drip line we may encounter nuclei for which the two-neutron emission is energetically possible, but can be hindered by different configurations of the parent and daughter nuclei.

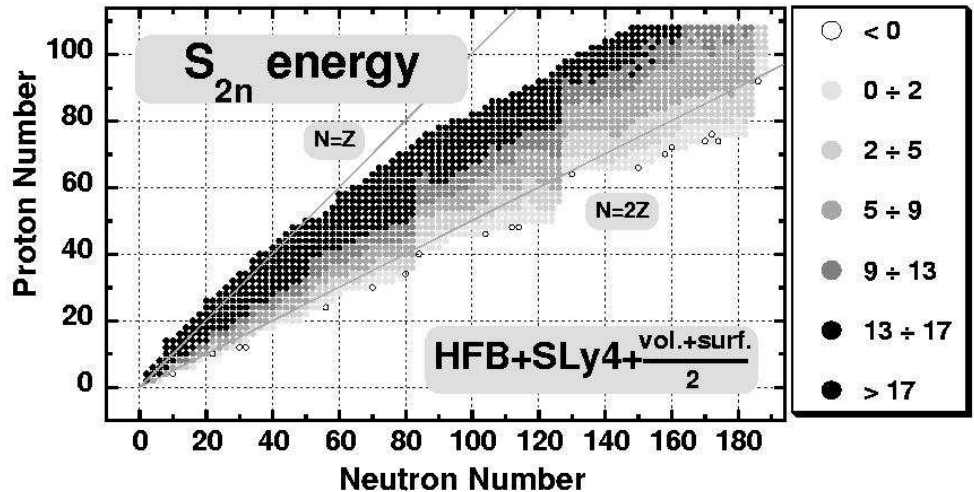

**FIGURE 5.** Two-neutron separation energies in even-even nuclei, obtained within the Skyrme-HFB method for the Skyrme SLy4 interaction and contact pairing force. From Ref. [79].

## NEUTRON-PROTON PAIRING COLLECTIVITY

Subject of the neutron-proton (np) pairing in nuclei has been studied since the early sixties, however, recently it witnessed an impressive renewal of interest which was due to increased experimental possibilities of studying medium heavy $N \approx Z$ nuclei. High-spin aspects of the np pairing are discussed in Ref. [10], therefore, here we concentrate on the question of collective ground-state correlations of this type. Despite the long history, the subject is plagued with contradicting opinions and results, and very often even the definitions of basic quantities are under debate. Without attempting a review of the whole issue, let us present here several important recent steps in this domain.

Question of the np pairing is very often discussed in connection to the so-called Wigner energy term in the mass formula, see Ref. [87] and references cited therein. Such a term represents an additional binding of $N \approx Z$ nuclei with respect to a *smooth-reference* dependence on $N$ and $Z$. The same additional binding is observed when the experimental masses are compared to *mean-field* mass predictions. It turns out that both, smooth part of experimental masses and mean-field masses, are in a perfect agreement with one another, and can be described by a term proportional to $(N-Z)^2=4T_z^2=4T^2$, where $T_z$ and $T$ are the ground-state values of the projection on the third axis and total isospin, respectively. On the other hand, the additional binding (the Wigner energy) is a function of $|N-Z|=2|T_z|=2T$, and therefore has a cusp at $N=Z$. Moreover, when added together both contributions to the binding energy give approximately a term proportional to $T(T+1)$, i.e., the numerical coefficients in front of the smooth term proportional to $T^2$, and the cusp proportional to $T$, are almost equal.

This simple observation has been known since many years [88], and recently was reanalyzed in Refs. [89, 90]. A controversy, however, arises at the level of interpretation of this result, namely, the authors of Refs. [89, 90] consider the $T(T+1)$ term as a simple symmetry energy that has nothing to do with pairing correlations, and hence claim that there is no experimental indication for the collective np pairing in nuclei. On the other hand, the authors of Refs. [87, 91] consider the mean-field term $T^2$ as representing the symmetry energy, and attribute the remaining contribution (proportional to $T$) to the $T=0$ np pairing correlations.

The dispute, however, is far from being an academic discussion triggered by a misunderstanding of definitions. Indeed, what is usually meant by correlations is the question whether an experiment that probes for a simultaneous presence of a neutron and a proton gives the probability higher or not than the product of probabilities to find a neutron and a proton in two independent experiments. Since the mean-field approximation gives *by definition* no correlations (in the sense of the above definition) it is reasonable to attribute all the effects, or contributions to energy, that are beyond the mean-field approximation, to correlations.

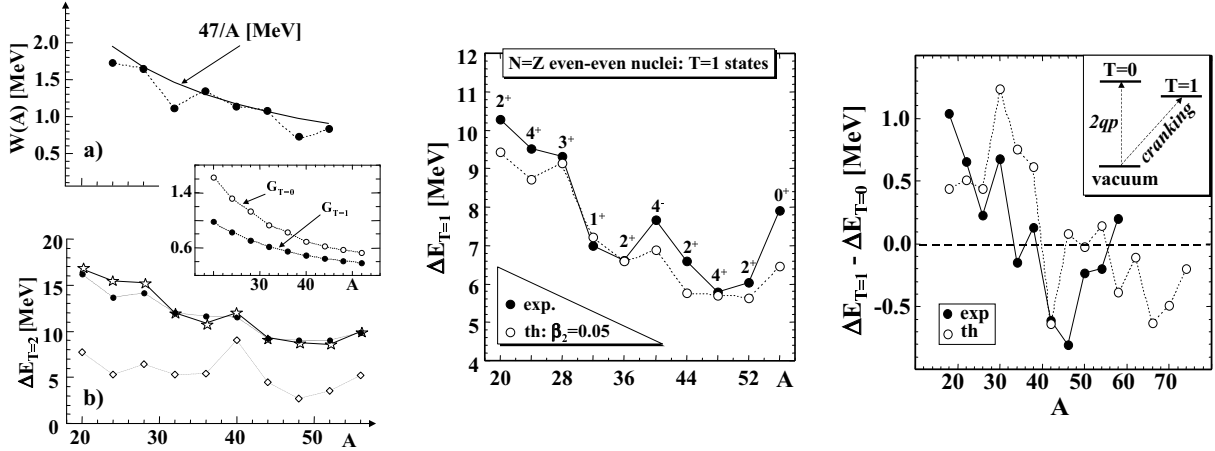

**FIGURE 6.** The Wigner energy (left top panel) and pairing coupling constants (left middle panel) together with the $T=0$, 1, and 2 excitation energies in $N=Z$ even-even and odd-odd nuclei. In the left bottom panel, experimental data (stars) are compared with the results of calculation with (full symbols) and without (open symbols) the $T=0$ np interaction included. From Ref. [92]

As was proved in Ref. [87], the full shell-model calculations in the $sd$ and $fp$ shells give the binding energies roughly proportional to $T(T+1)$, and hence they perfectly reproduce the Wigner energy. This result is, of course, not a trivial result of the isospin invariance of the interaction. Indeed, the isospin invariance requires that energies are functions of the total isospin $T$, and not necessarily functions of the quadratic invariant $T(T+1)$. [Similarly, not all rotationally-invariant Hamiltonians give rotational spectra proportional to $J(J+1)$.] The fact that the nuclear binding energies do behave as $T(T+1)$ is a non-trivial result of a specific structure of the nuclear (shell-model) interaction. In Ref. [87], it was also shown that the Wigner term disappears when all $T=0$ matrix elements are removed from the Hamiltonian. However, this observation does not constitute a proof that the $T=0$ pairing correlations are at the origin of the Wigner term. The shell-model calculations take into account all sorts of correlations, and it is rather difficult to disentangle and quantify the np pairing correlations in the final result. On the other hand, an approach that starts from the mean-field approximation and than adds correlations on top of it, is tailored towards such a quantification.

The first consistent description of several experimental facts observed in $N \approx Z$ nuclei was recently obtained [92] within the mean-field plus pairing approach. The authors were able to include all pairing channels simultaneously on top of the standard mean-filed solutions, and showed that the $T=0$ np pairing is essential to obtain agreement with several types of observables. Their results are summarized in Fig. 6. After an adjustment of the $T=0$ pairing coupling constant to the Wigner energy term (top left panel) they can reproduce (without any additional fitted parameters), (i) the $T=2$ states in even-even $N=Z$ nuclei (bottom left panel), (ii) the $T=1$ states in even-even $N=Z$ nuclei (middle panel), and (iii) the $T=1$ and $T=0$ states in odd-odd $N=Z$ nuclei (right panel). There are two key elements of their description, namely, the idea of the iso-cranking, i.e., a consistent application of the broken-symmetry treatment of nuclear excitations to the isospin channel, and (ii) identification of two-quasiparticle excitations among states of a given isospin, that is based on analyzing their time-reversal and iso-signature properties. Although these results are not a direct proof of existence of the collective np pairing correlations in nuclei, they are the best circumstantial evidence thereof, that is available to date.

# MAGNETIC ROTATION

Contrary to the standard image of nuclear rotation, where a deformed charge distribution rotates in the laboratory frame, a new type of rotational bands was recently discovered and interpreted as rotations of deformed distributions of currents, see reviews in Refs. [93, 94]. Net non-zero distributions of currents may occur in nuclei whenever proton and neutron angular momenta are misaligned at zero angular frequency, and then align when the angular frequency increases. This leads to a new way to build up high values of angular momenta, called the shears mechanism, in which the proton and neutron angular momenta align along a common axis. When the difference of proton and neutron angular momentum decreases, so decreases the magnetic moment, as illustrated in the right panel of Fig. 7, and hence the reduced magnetic dipole transition probabilities decrease with increasing angular momentum. During the whole alignment process the total angular momentum axis is tilted with respect to principal axes of the charge distribution, and hence the signature symmetry must be broken in the intrinsic frame of reference.

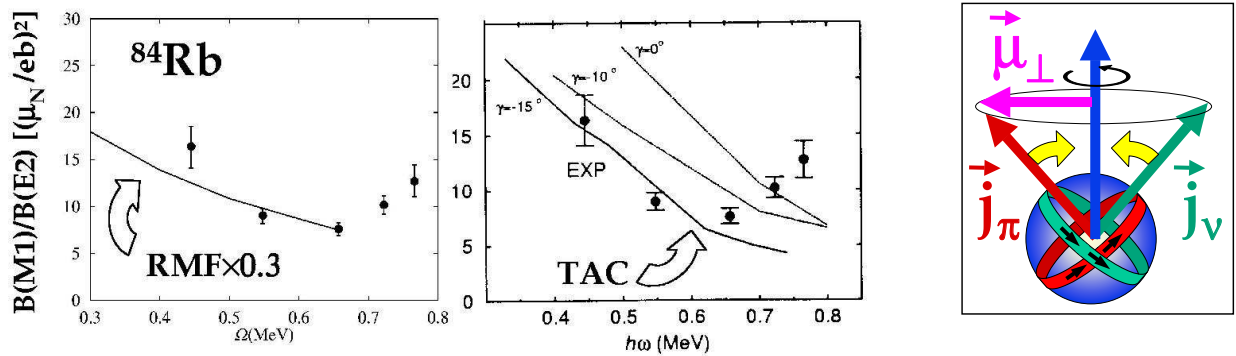

**FIGURE 7.** Schematic illustration of the shears mechanism in nuclei (right panel) together with the TAC (middle panel, From Ref. [95]) and RMF (left panel, From Ref. [96]) description of the magnetic band in $^{84}$Rb.

The shears bands, or the magnetic rotation phenomena, have already been discovered in several regions of nuclei, see, e.g., recent studies around $A \simeq 80$ [95], $A \simeq 110$ [97], $A \simeq 140$ [98, 99], and $A \simeq 200$ [100, 101, 102]. Most of such structures were found in light lead isotopes, but in principle, they can occur in any nucleus with the shell structure based on high-$j$ proton (neutron) holes coupled to high-$j$ neutron (proton) particles [93].

Up to now, the shears bands identified in the experiment were almost uniquely interpreted within the tilted-axis cranking (TAC) model proposed and developed by Stefan Frauendorf, see Ref. [103] and references cited therein. The model uses the standard phenomenological mean-field potential with pairing correlations, and correctly reproduces the main feature of the shears bands, namely, the decrease of $B(M1)$ with spin, see the example for $^{84}$Rb shown in the middle panel of Fig. 7. The only available self-consistent calculations of the shears mechanism were performed within the RMF approach [96], see the left panel of Fig. 7, and within the Skyrme-HF approach [104]. A more extensive self-consistent studies are required to elucidate the delicate balance between the standard (rotating charge) and new (rotating current) mechanisms of collective bands in weakly deformed nuclei. Moreover, only such studies can give us access to isovector time-odd terms in the rotating mean fields, which are neglected in the phenomenological mean-field potentials.

## PSEUDOSPIN SYMMETRY

The pseudospin symmetry has been introduced in nuclear structure physics many years ago [105, 106], in order to account for "unexpected" degeneracies in single-particle spectra. It has been attributed to an accidental cancelation between the spin-orbit field $\vec{\ell} \cdot \vec{s}$ and the $\vec{\ell}^2$ term in the Nilsson potential [107], the latter reflecting the fact that the nuclear single-particle field has a flat bottom due to the saturation of nuclear forces. Later, many similar degeneracies were also observed among rotational bands, see, e.g., the pair of bands in $^{187}$Os studied in Ref. [108] and references cited therein.

Recently, a very elegant and natural explanation was suggested [109, 110, 111], which is based on simple relativistic arguments and properties of the RMF Dirac equation for nucleons. Indeed, when the Dirac Hamiltonian is solved with the vector and scalar potentials, $V_V$ and $V_S$, respectively,

$$H = \vec{\alpha} \cdot \vec{p} + \beta(M + V_S) + V_V,$$ (3)

and when $V_V + V_S =$const, solutions can be classified according to a new SU(2) group, called the pseudospin group generated by $\vec{\tilde{s}} = U_p \vec{s} U_p$ for $U_p = 2(\vec{s} \cdot \vec{p})/p$. Then, the pseudo-orbital symmetry $\vec{\tilde{\ell}} = \vec{j} \cdot \vec{s}$ is also conserved, and the lower-component radial wave functions of the pseudospin partner states are identical to one another.

In finite nuclei, condition $V_V + V_S =$const is, of course, violated because the potentials have finite range, and hence, they must depend on the distance from the center of the nucleus. However, it turns out that in realistic situations the lower Dirac components of the pseudospin partners are still fairly close to one another, and their single-particle energies are almost degenerate. This is illustrated in Fig. 8 [112], where the upper (left panel) and lower (right panel) components of two bound states, $2s_{1/2}$ and $1d_{3/2}$, in $^{208}$Pb are plotted. These two states are the $\tilde{p}$ pseudospin partners corresponding to $\tilde{\ell}$=1, i.e., to the $1\tilde{p}_{1/2}$ and $1\tilde{p}_{3/2}$ orbitals, and their lower Dirac components are very similar indeed.

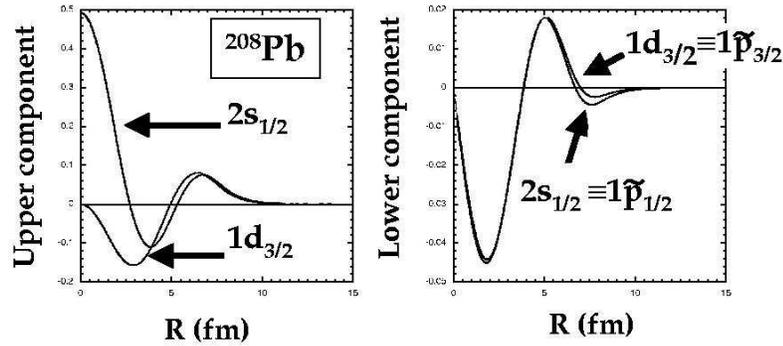

**FIGURE 8.** Upper (left panel) and lower (right panel) components of the Dirac wave functions for the $2s_{1/2}$ and $1d_{3/2}$ states in $^{208}$Pb. (From Ref. [112]. Reprinted from Nuclear Physics A, Vol 690, 2001, pp 41-51, J.N. Ginocchio, "A relativistic symmetry in nuclei: its...", Copyright (2001), with permission from Elsevier Science.)

Nevertheless, the utility of the idea of the pseudospin symmetry crucially depends on the symmetry breaking schemes in real nuclei, and this aspect of the proposed relativistic explanation requires further study, see the analyses presented in Refs. [113, 114] and references cited therein. In particular, an explanation of the fact that the pseudospin symmetry does not hold in light nuclei is still lacking (relativity arguments should hold irrespective of the number of particles). Similarly, it would be very interesting to see if the pseudospin symmetry still holds in neutron rich nuclei, where changes in the surface diffuseness may act against it [76].

## SHELL EFFECTS IN STRONG COULOMB FIELDS

Recent experimental discoveries and searches for new superheavy elements, see recent reviews in Refs. [115, 116], have triggered a substantial increase in the theoretical efforts to describe the structure and production of such systems, see recent Refs. [117, 118, 119, 120, 121, 122] and [123, 124, 125, 126, 127], respectively. Although many nuclides in the uncharted region of $Z$>112-114 are predicted to have substantial barriers against fission, they can be experimentally produced only with extremely low cross-sections. Stability and existence of superheavy nuclei are entirely due to the shell effects, and hence a delicate balance between extremely strong Coulomb field and nuclear mean-field potential must be self-consistently taken into account [128].

Due to the large numbers of protons and neutrons, densities of single-particle states in superheavy and hyperheavy nuclei are much larger than those in usual stable nuclei [121]. Hence, the magic shell gaps are significantly smaller, and the corresponding shell corrections shown in Fig. 9 are not-so-well localized around the doubly-magic nuclei. This observation [121] decreases the importance of predicting which is the next proton magic number after $Z$=82 [119]. Indeed, we may expect a fairly wide island of superheavy nuclei with tangible life times, and not a single or several long-living superheavy nuclides. However, precise estimates of life times require a better determination of nuclear effective forces and reaction mechanisms, when they are used for extrapolations to very exotic systems.

## PERSPECTIVES AND OUTLOOK

Focus of the present-day nuclear structure theory is on determination of nuclear effective interactions and/or energy density functionals. The term "effective" pertains here to the fact that they are used in a restricted phase space (shell model) or neglect correlations (mean field), and hence cannot be identical to interactions in the vacuum. Both types of methods are at present very well developed and operational, however, much more work is needed to find the best possible interactions to be used within them.

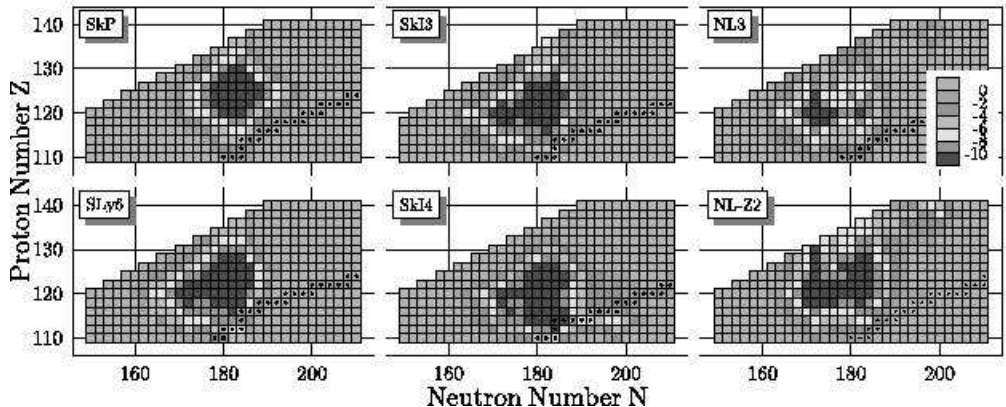

**FIGURE 9.** Shell corrections calculated from spherical self-consistent single-particle spectra for even-even nuclei around $Z=120$ and $N=180$. (From Ref. [121]. Reprinted from Physics Letters B, Vol 515, 2001, Page 42, M. Bender *et al*, "Shell stabilization of super- and hyperheavy nuclei without magic gaps", Copyright (2001), with permission from Elsevier Science.)

First-principle derivations of effective interactions are currently pursued with increased intensity, however, the task is difficult and requires some time to bear fruit. In the past such derivations were always complemented by phenomenological adjustment to experimental data. There is no doubt that such adjustments will remain the best source of recognizing the main features of interactions, and achieving a precise description. Therefore, future developments in nuclear structure theory critically depend on the availability of data that probe and enhance the most interesting and least known channels of effective interactions. The isospin properties are certainly of that kind, and we hope to get qualitatively new insight in this channel from the new generation rare-isotope facilities that are currently being constructed or planned throughout the world.

## ACKNOWLEDGMENTS

During preparation of this talk I have received suggestions and comments from over 40 of my colleagues and collaborators; I would like to thank them very much for their help. In particular, I would like to thank M. Bender, K. Bennaceur, S. Ćwiok, S. Frauendorf, S. Goriely, L. Grigorenko, M. Huyse, W. Nazarewicz, S. Pieper, M. Pearson, M. Riley, and K. Rykaczewski for sending me files with figures. This work was supported in part by the Polish Committee for Scientific Research (KBN) under Contract No. 5 P03B 014 21.